\documentclass[12pt]{article}%
\usepackage{amsmath}
\usepackage{amssymb}
\usepackage{amsfonts}
\usepackage{graphicx}%
\input epsf.tex
\newcommand{\be}{\begin{equation}}
\newcommand{\ee}{\end{equation}}
\newcommand{\bea}{\begin{eqnarray}}
\newcommand{\eea}{\end{eqnarray}}

\begin{document}
\bigskip\begin{titlepage}
\begin{flushright}
UUITP-09/06\\
hep-th/0606474
\end{flushright}
\vspace{1cm}
\begin{center}
{\Large\bf Transplanckian signatures in WMAP3?\\}
\end{center}
\vspace{3mm}
\begin{center}
{\large
Ulf H.\ Danielsson} \\
\vspace{5mm}
Institutionen f\"or Teoretisk Fysik, Uppsala Universitet, \\
Box 803, SE-751 08
Uppsala, Sweden
\vspace{5mm}
{\tt
ulf@teorfys.uu.se \\
}
\end{center}
\vspace{5mm}
\begin{center}
{\large \bf Abstract}
\end{center}
\noindent
In this note we investigate how a possible signal in the WMAP3 data of rapid oscillations in the primordial spectrum can be
accomodated into an effective model of transplanckian physics including back reaction. The results, if due to a real effect,
would indicate the presence of a low fundamental scale -- possibly the string scale -- around $2.2\cdot10^{-5}M_{pl}$.
\vfill
\begin{flushleft}
June 2006
\end{flushleft}
\end{titlepage}\newpage


\section{Introduction}

\bigskip

Recently, the third year of data from the WMAP satellite was released. The
high accuracy measurements provide an opportunity not only to test inflation
in general, but also to distinguish between specific models,
\cite{Spergel:2006hy}. One can even hope to obtain some important constraints
on models inspired by string theory.

A particularly exciting prospect is to look for qualitatively new effects due
to physics at or beyond the string or Planck scale. As have been argued in
many works -- the references [2-24] just represent a selected few -- there are
reasons to expect a characteristic signal consisting of a modulation in the
primordial spectrum with a periodicity determined by the slow roll parameters.
The magnitude of the effect is believed to be quite small but could
nevertheless be within reach of present or upcoming observatories.

In \cite{Spergel:2006hy} an analysis in search for such an effect was made but
no significant signal was found. Intriguingly, another analysis made in
\cite{Martin:2006rs} claim that there actually are some weak hints in the
present data. According to \cite{Martin:2006rs}, the indications have become
slightly stronger in light of the WMAP3 release compared to similar earlier
claims by the same authors, [29-31]. Even though the results are by no means
statistically significant, it is nevertheless an interesting and useful
exercise to see whether the data can be made compatible with the theoretical
expectations. If nothing else, it serves as an illustration of what kind of
constraints could be obtained in the event a real signal was discovered.

As noted in \cite{Martin:2006rs}, the parameters suggested by the data
correspond to a rapid oscillation with a surprisingly large amplitude. It is
important to note that the claimed oscillations in amplitude are periodic in
the logarithm of the scale of the CMBR fluctuations ; just as expected if due
to transplanckian physics. This is related to the underlying assumption that
there are new physics associated with a fixed energy scale. Other possible
sources of modifications in the primordial spectrum -- such as features in the
inflaton potential -- are typically associated with physics taking place at
some definite moment in time, and give rise to other types of oscillations,
\cite{Burgess:2002ub}\cite{Burgess:2003zw}\cite{Kaloper:2003nv}.

Given the strength of the effect it is a valid concern whether it is at all
consistent with a possible problem of back reaction on the geometry. After
all, the oscillations are supposed to be caused by a nontrivial vacuum for the
inflaton field that also would be expected to contribute to the energy density
and change the way the universe expands. As shown in
\cite{Brandenberger:2004kx} and \cite{Danielsson:2004xw}, however, even a
large back reaction will not necessarily destroy the inflationory phase. The
main purpose of this note is to address this  issue in the context of the
analysis made in \cite{Martin:2006rs}. We will make use of a formalism
developed in \cite{Danielsson:2004xw} and \cite{Danielsson:2005cc} where the
effect of back reaction is incorporated in a self consistent way. We will show
that the back reaction is under control and fully consistent with inflation,
with a slow roll found to be completely dominated by the vacuum energy given
the parameters suggested by \cite{Martin:2006rs}. To be precise, we find
evidence for a fundamental scale around $2.2\cdot10^{-5}M_{pl}$, where
$M_{pl}\sim2.4\cdot10^{18}\mathrm{GeV}$ is the reduced Planck mass.

The outline of the paper is as follows. We begin in section two with a summary
of the phenomenological model we are using, including back reaction. We then
proceed with an analysis of the WMAP3 data as given by \cite{Martin:2006rs},
and end with some conclusions.

\bigskip

\section{A model of transplanckian physics}

\bigskip

The main idea, which consequences we want to explore, is that physics beyond
the string or Planck scale is magnified through the expansion of the universe,
and affects phenomena at lower energies such as the fluctuations of the CMBR.
We will model possible new physics with a choice of vacuum different from the
usual Bunch-Davies vacuum, where we assume a Bogolubov mixing linear in
$\frac{H}{\Lambda}$ with the only dependence on scale being through the Hubble
constant $H$. $\Lambda$ is the energy scale of the new physics which could be
the string scale or the Planck scale. We assume that all this new physics can
effectively be encoded in the choice of vacuum.

As argued in \cite{Danielsson:2005cc}, the effects propagating down to low
energy will be of two types: a modulation of the CMBR spectrum and a back
reaction on the expansion of the universe. Below we will review what the
consequences are.

\bigskip

\subsection{Effects on the CMBR}

\bigskip

According to the analysis of \ \cite{Danielsson:2002kx}, given a Bogolubov
mixing as described above, the typical effect to be expected on the primordial
spectrum is of the form%

\begin{equation}
P(k)=\left(  \frac{H}{\overset{\cdot}{\phi}}\right)  ^{2}\left(  \frac{H}%
{2\pi}\right)  ^{2}\left(  1-\frac{H}{\Lambda}\sin\left(  \frac{2\Lambda}%
{H}\right)  \right)  ,
\end{equation}
where we note a characteristic, relative amplitude of the correction given by
$\frac{H}{\Lambda}$, and a modulation sensitively dependending on how
$\frac{\Lambda}{H}$ changes with $k$. The claim is that whatever the nature of
the high energy physics really is, a modulated spectrum of this form is what
we should naturally expect. In \cite{Bergstrom:2002yd} one can find an early
discussion of the phenomenological relevance of the effect, and how the
magnitude is related to the characteristic parameters describing the
inflationary phase. Using the standard slow roll approximation, where an
important parameter is
\begin{equation}
\varepsilon=\frac{M_{pl}^{2}}{2}\left(  \frac{V^{\prime}}{V}\right)  ^{2},
\end{equation}
with initial conditions imposed at some fundamental scale $\Lambda=\gamma
M_{pl}$, it is found that
\begin{equation}
\frac{\Delta k}{k}\sim\frac{\pi H}{\varepsilon\Lambda}\sim1.3\cdot10^{-3}%
\frac{1}{\gamma\sqrt{\varepsilon}},
\end{equation}
and
\begin{equation}
\frac{H}{\Lambda}\sim4\cdot10^{-4}\frac{\sqrt{\varepsilon}}{\gamma
}.\label{eq:effect}%
\end{equation}
These two relations are the key to estimating the expected magnitude of the
effect. For instance, with a string scale a couple of order of magnitudes
below the Planck scale, and a slow roll parameter $\varepsilon\sim10^{-2}$, we
find an amplitude of $\frac{H}{\Lambda}\sim10^{-2}$ -- comparable with cosmic
variance -- and a periodicity given by $\frac{\Delta k}{k}\sim\mathcal{O}%
\left(  1\right)  $. The data suggested by WMAP3, however, need a
generalization of the analysis to be accommodated which we turn to below.

\bigskip

\subsection{Back reaction}

\bigskip

As explained in the introduction, the presence of a nontrivial vacuum,
motivated by the presence of unknown high energy physics, raises the issue of
backreaction. Focusing on the contribution to the vacuum energy coming form
the non-standard vacuum, as compared with the Bunch-Davies vacuum, one finds
an additional energy density naively given by $\rho_{\Lambda}\sim\Lambda
^{2}H^{2}$. To lowest order, as long as $\Lambda\ll M_{p}$, we can ignore this
contribution as was concluded in \cite{Tanaka:2000jw}. In
\cite{Danielsson:2004xw} and \cite{Danielsson:2005cc}, however, the discussion
was taken a step further and it was noted that the presence of the background
energy will change the effective slow roll parameters. In fact, in our attempt
to match the possible WMAP3-effect, we will find ourselves in a situation
where it is the non-standard physics that dominates the slow roll of the
Hubble constant.

We denote the relevant slow roll parameter by $\varepsilon$ and define it
through%
\begin{equation}
\varepsilon=\frac{\dot{H}}{H^{2}}.
\end{equation}
In addition, we still assume a rolling inflaton in the background whose main
role is to end inflation. That is, it is initially governed by a slow roll
parameter according to%
\begin{equation}
\varepsilon_{\inf}=\frac{\dot{\phi}^{2}}{2M_{pl}^{2}H^{2}},\label{einf}%
\end{equation}
where $\phi$ is a canonically normalized inflaton. It is $\varepsilon_{\inf}$
which, in the usual way, determines the relative amplitude between the
dominating scalar modes and the tensor modes. With no back reaction from the
vacuum we would have had $\varepsilon=\varepsilon_{\inf}$, while we here are
interested in the case where $\varepsilon_{\inf}\ll\varepsilon\ll1$, during
the era when the fluctuations relevant for the CMBR are generated. Inflation
ends when $\varepsilon_{\inf}$ has evolved to become of order one. As
explained in \cite{Danielsson:2005cc}, this leads to a decoupling of the
expressions for the amplitude and the period according to
\begin{equation}
\frac{H}{\Lambda}\sim4\cdot10^{-4}\frac{\sqrt{\varepsilon_{\inf}}}{\gamma},
\end{equation}
and%
\begin{equation}
\frac{\Delta k}{k}\sim\frac{\pi H}{\varepsilon\Lambda}\sim1.3\cdot10^{-3}%
\frac{\sqrt{\varepsilon_{\inf}}}{\gamma\varepsilon}.\label{eq:period}%
\end{equation}

Let us now proceed with an estimate of $\varepsilon$, following
\cite{Danielsson:2004xw}. The above estimate of the energy density is not good
enough when we want to find an expression for $\varepsilon$. What we need to
do is to take into account that $H$ will be changing with time, i.e. decrease.
Modes with low momenta were created at earlier times when the value of $H$
were larger, and there will be an enhancement in the way these modes
contribute to the energy density. We therefore find an energy density given
by
\begin{equation}
\rho_{\Lambda}\left(  a\right)  =\frac{1}{2\pi^{2}}\int_{\varepsilon}%
^{\Lambda}dpp^{3}\frac{H^{2}\left(  \frac{ap}{\Lambda}\right)  }{\Lambda^{2}%
}=\frac{1}{2\pi^{2}}\frac{\Lambda^{2}}{a^{4}}\int_{a_{i}}^{a}daa^{3}%
H^{2}\left(  a\right)  ,
\end{equation}
where we have introduced a low energy cutoff corresponding to the energy at
the time of observation of modes that started out at $\Lambda$ at some
arbitrary initial scale factor $a_{i}$. If we take a derivative of the energy
density with respect to the scale factor and use $\frac{d}{da}=\frac{1}%
{aH}\frac{d}{dt}$, we find%
\begin{equation}
\dot{\rho}_{\Lambda}+4H\rho_{\Lambda}=\frac{1}{2\pi^{2}}\Lambda^{2}%
H^{3},\label{eq:contQ}%
\end{equation}
and we conclude that we must introduce a source term in the Friedmann
equations. It was found in \cite{Danielsson:2004xw} that the evolution is
governed by%
\begin{equation}
\frac{d}{da}\left(  a^{5}HH^{\prime}\right)  =-\frac{\Lambda^{2}}{3\pi
^{2}M_{pl}^{2}}a^{3}H^{2}-\frac{1}{2M_{pl}^{2}}\frac{d}{da}\left(
a^{4}\left(  aH\phi^{\prime}\right)  ^{2}\right)  ,
\end{equation}
where we let $%
\acute{}%
=\frac{d}{da}$. The first term on the right hand side is due to the presence
of the non-standard vacuum, while the second term is due to the presence of
the inflaton potential. In a situation where the first term dominates, we find
a slow roll governed by
\begin{equation}
\varepsilon=\frac{\gamma^{2}}{12\pi^{2}},
\end{equation}
for small $\gamma=\frac{\Lambda}{M_{pl}}$.\footnote{In this paper we
consistently use the reduced Planck mass $M_{pl}=1/\sqrt{8\pi G}\sim
2.4\cdot10^{18}\mathrm{GeV}$. This should be kept in mind when comparing with
the results in \cite{Danielsson:2005cc}.} In the standard case, with no vacuum
contribution, the only non-vanishing term is the second one, leading to a slow
roll governed by (\ref{einf}).

There are two ways to further generalize the model, one of which played an
important role in \cite{Danielsson:2005cc}. While it is the slow roll of the
inflaton that controls the overall amplitude of the primordial spectrum, one
could easily imagin that there are more fields in the non-standard vacuum.
These would also contribute to the back reaction and enhance $\varepsilon$ by
a factor $n$, where $n$ is the number of participating fields. Another
generalization was advocated in \cite{Armendariz-Picon:2003gd}, where it was
argued that one should allow for a free numerical factor in the Bogolubov
coefficient and generalize $\frac{H}{\Lambda}\rightarrow x\frac{H}{\Lambda}$.
While this makes the model less natural, it is a phenomenological parameter
which observations will need to tell us the value of, and it is in fact
necessary in order to make sense of the hints in WMAP3. With these new
parameters we find%
\begin{equation}
\varepsilon=\frac{nx^{2}\gamma^{2}}{12\pi^{2}},\label{eq:ebak}%
\end{equation}
and that the amplitude of the oscillation is controlled by%
\begin{equation}
x\frac{H}{\Lambda}\sim4\cdot10^{-4}x\frac{\sqrt{\varepsilon_{\inf}}}{\gamma
}.\label{eq:kbak}%
\end{equation}

\bigskip

\section{Application to WMAP3 and conclusions}

\bigskip

We are now ready to apply our phenomenological model to the actual data
provided by WMAP3 as given in \cite{Martin:2006rs}. According to the analysis,
the values giving the best fit to the data are found to be%
\begin{equation}
x\frac{H}{\Lambda}\sim0.27,\ \varepsilon\sim2.1\cdot10^{-3},\ \frac{\Delta
k}{k}\sim0.018,\ \mathrm{and}\ x\sim22500.
\end{equation}
We note that the amplitude of the claimed oscillations is quite large and
implies an unnaturally large value for the dimensionless number $x$%
.\footnote{In fact, there does not seem to be any suppression of the amplitude
due to the ratio between the Hubble scale and the fundamental scale.} We also
note that the oscillations are very rapid -- tens of oscillations imposed on
every acoustic peak.

Making use of the above values and the key equations (\ref{eq:period}),
(\ref{eq:ebak}) and (\ref{eq:kbak}), we can now easily estimate the parameters
in our model which match the data. The results we find are%
\begin{align}
\gamma &  \sim2.2\cdot10^{-5}\\
\varepsilon_{\inf} &  \sim5.3\cdot10^{-11},
\end{align}
where we have assumed $n=1$; with more participating fields the values become
even lower. We note that our ansatz is self consistent with $\varepsilon
_{\inf}\ll\varepsilon\ll1$, and that we find a very low string scale and even
lower Hubble scale.

As we have seen there are two extremes to consider when we match the data. On
the one hand we have the possibility considered in, e.g.
\cite{Bergstrom:2002yd}, where the inflationary parameters are dominated by
the inflaton and the back reaction due to the vacuum energy is very small. In
this case the usual relations between the slope of the spectrum, the relative
magnitude of scalar and tensor modes, and the slow roll parameter
$\varepsilon$ apply. In the other extreme -- relevant for the WMAP3 data --
the vacuum energy dominates and the slow roll is directly linked to the
fundamental scale. The inflaton, on the other hand, is almost locked and
changes only very slowly. As a consequence the tensor modes are much further
suppressed compared with the scalar ones than what one first would expect, and
we find ourselves in a situation similar to hybrid inflation. Eventually, the
inflaton must enter into a regime where its potential starts to change rapidly
and finally make inflation come to an end.

In this note we have argued that the effects discussed in \cite{Martin:2006rs}%
, if real, can be accommodated in an effective transplanckian model with a low
string scale. It is interesting to note how the model unambiguously (up to the
number of fields $n$) constrain the fundamental scale and also tell us how the
inflaton rolls even though its contribution is subdominant. While highly
speculative, this exercise nevertheless shows how new observational data can
be used to constrain physics relevant for string theory and quantum gravity.

\bigskip

\section*{Acknowledgments}

The work was supported by the Swedish Research Council (VR).

\end{document}